\documentclass[12pt,preprint]{aastex}

\slugcomment{}

\shorttitle{A Flare of AE Aquarii Observed with XMM-Newton}
\shortauthors{Choi and Dotani}

\begin{document}

\title{A Flare of AE Aquarii Observed with XMM-Newton\footnote{An ESA
science mission with instruments and contributions directly funded by
ESA Member States and NASA}} 

\author{Chul-Sung Choi\altaffilmark{*} and Tadayasu Dotani\altaffilmark{**}}
\email{cschoi@kasi.re.kr and dotani@astro.isas.jaxa.jp}

\altaffiltext{*}{Korea Astronomy and Space Science Institute, 36-1 Hwaam,
Yusong, Taejon 305-348}
\altaffiltext{**}{Institute of Space and Astronautical Science, Japan
Aerospace Exploration Agency}


\begin{abstract}

We present the results of analyzing the XMM-Newton data obtained in
2001 November 7 - 8.
A flare is observed simultaneously in X-ray and UV together
with a quiescence. 
We find that during the flare event X-ray flux varies with UV with
no significant time lag, indicating a close correlation of flux
variation for X-ray and UV flares.
An upper limit of the lag is estimated to be $\sim 1$~min.
From a timing analysis for X-ray data, we find that both pulsed
and unpulsed flux increase clearly as the flare advances in the entire
energy band 0.15 - 10~keV.
The net increase of pulsed flux to the quiescence is, however, small
and corresponds to about 3 $-$ 4\% of the increase in unpulsed flux, 
confirming that a flux variation of flare in AE Aqr is dominated by
unpulsed X-rays.
A spectral analysis reveals that the energy spectrum is similar to
that of the quiescence at the beginning of the flare, but the
spectrum becomes harder as the flare advances.
Based on these results, we discuss the current issues that 
need to be clarified, e.g., the possible flaring site and the mass 
accretion problem of the white dwarf.
We also discuss the flare properties obtained in this study.

\end{abstract}

\keywords{binaries: close ---
stars: individual(\objectname{AE Aquarii}) ---
stars: cataclysmic variables --- X-rays: binaries}

\section{Introduction}

AE Aqr is an unusual nova-like magnetic cataclysmic variable (mCV).
It is known as a non-eclipsing close binary system with an orbital period
of P$_{\rm orb} \approx 9.88$~hr \citep[for a review see, e.g.,][]{war95}.
The primary is a magnetized white dwarf rotating at a period of
P$_{\rm spin} \approx 33.08$~s and the secondary (or the companion star)
is known to be a red dwarf of spectral type K3$-$K5.
It has been widely believed that the secondary transfers matter to the 
primary through a Roche-lobe overflow process \citep[e.g.,][]{cas96}, 
but H$\alpha$ Doppler tomograms of AE Aqr have revealed no evidence of an 
accretion disk \citep[e.g.,][]{wyn97, wel98}. 

Pulsations or modulations at the spin period have been detected from optical
to X-ray, but not in radio.
An optical/UV pulse profile exhibits sinusoidal double peaks, where the 
two peaks are separated by 0.5 in phase and their amplitudes are unequal
\citep[e.g.,][]{era94}.
On the other hand, an X-ray pulse profile exhibits a sinusoidal single 
peak \citep[e.g.,][]{choi99}.
It is widely believed that the pulsed radiations of optical/UV are 
produced at the magnetic pole areas of the rotating white dwarf, but
no consensus yet reached on the production mechanism of pulsed X-rays.
\citet{choi99} and \citet{choi00} suggested that the pulsed X-rays are 
produced by an accreting plasma onto the magnetic poles.
However, \citet{ikh01} argued that, under a strong $\lq$propeller' 
effect, matter overflowed from the Roche-lobe cannot reach the magnetic poles.
It is therefore necessary to clarify how the X-ray pulsations are produced.

AE Aqr is known to switch between the flaring and the quiescent 
states irregularly.
This is one of the peculiar properties that distinguishes this object
from other mCVs.
Various flares of different profiles and durations
($\sim$~a few hr) have been observed in radio, optical, UV, and X-ray
\citep[e.g.,][]{pat79, bas88, par89, era96, choi99}.
During optical flares, continuum flux often increases by a factor of 
several compared to the quiescent states.
According to the reports on simultaneous observations 
\citep[e.g.,][]{aba95, era96}, there is a good correlation between
optical and UV flares but not between optical and radio flares.
\citet{osb95} presented simultaneously obtained X-ray and optical
data that give us a hint of correlation, but their data seem
to be insufficient to examine the correlation of the flares 
because of a relatively short coverage of the X-ray data.
It is therefore somewhat ambiguous if X-ray flares indeed correlate
with optical/UV flares.

Nature of the X-ray flares of AE Aqr is not yet understood well.
The main results of ROSAT observations are: a pulse amplitude tends to
increase during the flares \citep{osb95},
and a flare spectrum can be fitted by only a two-temperature thin thermal
plasma emission model \citep{cla95}.
ASCA observation has revealed the following facts \citep[]{cd99, choi99}:
(1) an X-ray flux variation is dominated by unpulsed X-ray;
(2) a duration or a time scale of the flare in X-ray is comparable with
that of the optical/UV flares;
(3) an X-ray spectrum is very soft ($kT\sim$~3~keV) unlike the spectra of
other mCVs ($kT\sim$~a few tens keV);
(4) there is no significant difference between quiescence and flare spectra
(both are well fitted by a two-temperature thin thermal plasma emission model).

Although many flares have been reported for AE Aqr in various wave bands,
there is still a lack of consensus on the flaring site.
Furthermore it is poorly understood whether the flares in AE Aqr have
uniform properties or not.
Our present study is motivated by this situation.
In this paper, we concentrate on X-ray data and extract results associated 
with the current issues that need to be clarified.

\section{Instruments and Data}

AE Aqr was observed with XMM-Newton in 2001 November 7 through 8 
(Obs-Id 0111180201).
We acquired the data from the HEASARC archival data center and
reprocessed the data through the pipeline process with updated 
Current Calibration Files, using the XMM-Newton Science Analysis
Software SAS v5.4.1.
After this reprocessing, we extracted good event data by applying 
rather conservative screening criteria to the pipeline products of 
EPIC PN and EPIC MOS (e.g., $\lq$FLAG == 0' and $\lq$PATTERN $<$= 4'
in SAS).

XMM-Newton (Jansen et al. 2001) has three focal plane imaging spectrometers
with moderate spectral resolution: one European Photon Imaging Camera (EPIC)
PN detector \citep{str01} and two EPIC MOS detectors 
\citep[MOS1~\&~2,][]{tur01}. 
In addition, XMM-Newton carries one Optical Monitor \citep[OM,][]{mas01} 
and two identical Reflection Grating Spectrometers with a high spectral
resolution \citep[RGS1~\&~2,][]{her01}. 

During the observations of AE Aqr, PN was operated in imaging and full 
frame window modes, while MOS1~\&~2 were operated in imaging and large
window modes (all the EPIC instruments used medium filters).
Meanwhile, OM was operated in fast mode and used UVW2 filter
which is effective in wavelengths 2050~\AA $-$ 2450~\AA.
In this study we focus on the EPIC and the OM data.
The source events are extracted from a circular region of radius
45\arcsec\ for PN data which is centered at the position of AE Aqr,
and that of radius 135\arcsec\ for MOS1 and MOS2 data.
Time resolution of PN and MOS data are 73~ms and 0.9~s, respectively.

\section{Analysis of the Light Curve and the Flare}

\subsection{X-ray and UV Light Curves}

Figure~1 displays the XMM-Newton light curve of AE Aqr. We note that a
small background is not subtracted from the light curve.
An average MOS1 count-rate in the segment R1 is 1.24 $\pm$ 0.01 counts s$^{-1}$,
whereas an average count-rate of the background estimated from an annular region
of the same chip (inner and outer radii are 137\arcsec\ and 152\arcsec\ ,
respectively) is 0.09 $\pm$ 0.02 counts s$^{-1}$, which is scaled to the size
of the source region.
This indicates that the flare is not related with an instrumental background
flare. We show the light curve of the background data in Figure~2 for reference.

The orbital phase at the top of the figure~1 is calculated from the
ephemeris and orbital period presented by \citet{cas96}, where phase
0.0 corresponds to the superior conjunction of the white dwarf. 
We cross-checked the
calculation using other ephemerides and orbital periods presented by,
e.g., \citet{wel93} and \citet{rb94}, but found no significant difference. 
We here correct the orbital phase coverage of the ASCA light curve 
presented by \citet{choi99}, resulting in phases in the range 0.5 - 2.8.

As seen in Figure~1, a flare is simultaneously observed by EPIC and
OM detectors.
To understand how the X-ray properties evolved, we divide the light 
curve into five segments, $\lq$R0' to $\lq$R4'.
The X-ray light curve exhibits the following characteristics in
time variation (see Figure 1a and 1b):
(1) the X-ray flux decreases slightly in R0\footnote{There are no PN data
in this segment because the instrument was operated to have a different 
exposure time.} which lasts for $\approx 39.2$~min;
no distinct indications of an eclipse is noticed (see Figure 1b); 
(2) the flux variation is relatively small in R1 for $\approx 100$~min; 
therefore, this segment is regarded as a quiescent state;
(3) the flux increases steadily for $\approx 27.4$~min in R2 up to 3 times 
the average flux of the quiescence (an increasing state of flare);
(4) the increased high flux level is sustained throughout R3 for 
$\approx 66.7$~min (a flaring state) and there are many flickerings, i.e.,
small peaks having timescales of a few minutes and having relative amplitudes
of a few times 10\%;
(5) the flux again decreases for $\approx 34$~min in R4 (a decaying state 
of flare), after undergoing a large variation as compared with the
flickerings.
The light curve in R0 is clearly lacking in the evidence of an eclipse.
This confirms that AE Aqr is a non-eclipsing binary system. 
We note that the flickerings are more frequent during the flare, 
R2 $-$ R4, than the quiescence. 

Figure~1c displays the OM light curve, where we combine 5 pieces of OM 
data into one.
This figure exhibits that UV varies together with X-ray throughout the
entire phase.
We confirm that all of the X-ray variations discussed above are also seen
in the OM curve: the quiescence, the flare, and the flickerings.
According to visual inspection, there seems to be no
significant time lag between UV and X-ray light curves.
To confirm this, we pick up 4 flickering profiles which appear 
predominantly both in OM and MOS1 light curve, marked with arrows 
in Figure~1b \& 1c, and fit a Gaussian function\footnote{We adopted 
a simple Gaussian model for this fit and allow a Gaussian width to be
free for both UV and X-ray data.} to the profiles.
The measured time lags are 39~s~$\pm$~19~s, 32~s~$\pm$~14~s, 
17~s~$\pm$~10~s, and 0~s~$\pm$~10~s, respectively, in time order (the X-ray 
flickerings precede the UV flickerings).
There might be a hint of a positive time lag but the lag is at most 
2$\sigma$ level. Thus we conclude that the time lag is not statistically
significant and its upper limit is $\sim 1$~min.

\subsection{A Morphological Similarity of X-ray and UV Flares}

It is noteworthy that the flare light curve of EPIC (or OM) is very similar
to that of the UV flare presented by van Paradijs et al. (1989, see the 
left panel of Figure~2 in the paper, particularly the latter part).
It is also interesting to note that the ASCA flare presented by 
Choi et al. (1999) is very similar to the UV flare of 
Eracleous et al. (1994), although both data are rather sparse
(see, e.g., Figure~2 of Eracleous et al.; of the two flares the
latter is particularly consistent with the ASCA one).
Another interesting point is that the UV flares tend to
occur around the orbital phases at which the X-ray flares occur.
For example, the UV flare of Eracleous et al. covers the orbital
phases of $\Phi_{\rm orb} \sim$ 1.2 $-$ 1.5 and the UV flare 
of van Paradijs et al. covers $\Phi_{\rm orb} \sim$ 0.3 $-$ 0.4.
The latter coverage is similar to the flare of the present paper,
R2 $-$ R4, $\Phi_{\rm orb} = 0.25 -0.45$.

We conjecture from these results that the coincidence of EPIC 
and OM flare is not by chance.
The similarities of the flare light curve and the orbital phase 
coverages imply that a certain type of flare tends to occur 
preferentially at a certain orbital phase.

\section{Timing Analysis}

Here we investigate the behavior of pulsing and unpulsing X-rays
in connection with the light curve variation.
For this purpose, we searched the best period using an epoch folding
technique.
The pulse period was assumed to be constant throughout the observation.
We determined the best period, which gave the maximum $\chi^2$-value, 
by fitting a Gaussian function to the $\chi^2$-plot
as a function of the trial period.
This gives the best period of $P_{\rm s} = 33.067 \pm 0.005$~s, where
the uncertainty is calculated by the method presented by \citet{lar96}.
We used this period to examine the change of the pulse profile with
the development of the flare.
For this purpose, we subdivided the segments R1 $-$ R4 into smaller 
ones to avoid a large flux change within a subsegment, which might 
affect the pulse profile.
We then folded the data within the subsegment to obtain the pulse profile.
Pulse parameters (e.g., pulsed flux and pulse fraction) were calculated
through a model fit to the extracted profiles.

Figure 3 displays the pulse profiles (crosses) folded at $P_{\rm s}$
and the best-fit model function (solid curves) for each subsegment.
We adopted a sinusoidal model, i.e. a sine curve plus a constant
(see the table comments of Table~1 for the formula), to fit the
pulse profiles.
The best-fit parameters are listed in Table~1, where the pulse
amplitude is defined as the difference of the count rates at the
pulse minimum and the maximum, and the unpulsed flux as the count
rate at the pulse minimum.
As seen in Figure~3, all the pulse profiles exhibit a single
broad peak, which is different from the optical/UV pulse profile.
We checked the profiles by changing the number of phase bins, but
could not find any hints of an interpulse or a secondary peak.

Figure~4 shows how the pulse parameters change with the development of
the flare.
Through a careful inspection of this figure and the values in Table~1,
we find the followings:

\begin{enumerate}

\item The unpulsed flux (Figure~4b) increases by a factor of 5 in R4-1
compared with its quiescence value in R1-1 and its time history is
almost the same as that of the light curve. 
This confirms that the light curve variation is dominated by
unpulsed X-rays.

\item The pulse-amplitude variation (Figure~4c) is different from the
previous results by \citet{osb95} and \cite{choi99}. For example,
Osborne et al. found the amplitude to increase during a large flare only,
and Choi et al. found no significant difference in the amplitudes between
quiescence and flare (see also \S 6.3).
However, our present result clearly shows that the amplitude increases
from quiescence to flare.

\item The pulsed flux, i.e. half of the pulse amplitude, 
is larger by a factor of 2 in R4-1 than that in R1-1.
However, the net increase in the flux is small and corresponds to
only 3 $-$ 4\% of the increase of unpulsed flux.
It is interesting that the pulsed flux changes largely even
during quiescence: it increases by $\sim$75\% from R1-1 to R1-3,
while the unpulsed flux increases only by $\sim$35\% (see Table~1).

\item The pulse fraction (Figure~4d) varies in proportion to the
variation of total flux for the quiescence but it is inversely
proportional for the flare. This means that the variation of pulsed
flux is relatively large during the quiescence.

\end{enumerate}

From these results, we consider that, if the flare is caused by
an enhanced mass supply from the companion, most of the transferred
matter does not accrete onto the white dwarf surface.
This indicates that a propeller is operating in this system.
Nonetheless, a small amount of the matter must be accreted onto the
surface in order to explain the fact that pulsed flux exists
persistently through the observation.
As explained above, presence of the pulsed flux and its
increase are clear during the quiescence.

\section{Spectral Analysis}

In this section, we explore how the flux variation of the flare causes
the spectral variation using a model-independent method.
Then, we cross-check the result by performing a model fit to each
segment's spectrum.
Finally, we analyze an energy-resolved pulse profiles to diagnose a
spectral property for pulsed X-ray.

\subsection{PHA Ratios}

In order to study a relative spectral variation, we calculate
pulse-height$-$to$-$amplitude (PHA) ratios\footnote{That is, the
ratios of the count rates between two spectra as a function of X-ray 
energy.}
for R2 - R4 data, taking R1 spectrum as a reference.
Figures~5a to 5c show the ratios for PN data and Figure~5d shows
the ratios for the ASCA data presented by Choi et al. (1999).
We find that the ratios in Figure~5b and 5c are quite similar
to each other while the ratios in Figure~5a are comparable with that of
Figure~5d.

The ratios in Figure~5a are almost flat, indicating that the R2 spectrum
is very similar to the quiescence spectrum except for normalization.
On the other hand, the ratios in Figure~5b \&~5c are almost flat below
$\approx 0.5$~keV, increase steeply from 0.5~keV to 2.0~keV, and show
large scatters at higher energies.
This behavior suggests that both R3~\&~R4 spectra are similar to the
quiescence spectrum below $\approx 0.5$~keV; they are relatively harder
between 0.5~keV and 2.0~keV.
The large scatters at higher energies might be due to the sharp
structures, e.g. emission lines, in the energy spectrum.
These changes of the PHA-ratios suggest that plural emission
components are involved in the X-ray emission.
In the next subsection, we study the spectral changes through
model fitting using a multi-component model of optically
thin hot plasma.

\subsection{Spectral Fitting}

Recently, \citet{itoh04}, who analyzed the same data as ours,
reported that a 4-temperature $\lq$VMEKAL' model is required to 
fit the energy spectrum extracted from the whole EPIC data.
We adopt the same model in the present analysis because we confirmed
their spectral-fit result with our own data that cover the same energy range.
As shown in Table~2, we fit the spectrum of each segment with 
the 4-temperature model.
In this fit, we fixed the elemental abundances to the values obtained by
them and increased the number of spectral component step by step as
they did. 
Figure~6 displays a sample fit.

We obtain the following results from inspecting the fit values:
(1) all the obtained temperature values for R1 \& R2 spectra, $kT_1$ - $kT_4$,
are almost identical (R3 \& R4 spectra show similar result too);
(2) throughout the observation, the lowest-temperature values, $kT_1$, 
are almost constant together with their normalizations, which suggests that
this component exists persistently in AE Aqr system and is little affected
by a flaring activities (an unabsorbed flux of this component for the 
quiescence is calculated to be $8.4 \times 10^{-13}$ ergs cm$^{-2}$ s$^{-1}$
in the energy range of 0.15 - 10~keV);
(3) the highest-temperature values, $kT_4$, are much higher
for R3 \& R4 spectra, indicating that the energy spectrum
becomes hardest near the peak of the flare.

As mentioned above, the ratios of Figure 5d is similar to that
of Figure 5a rather than the ratios in Figure 5b \& 5c, suggesting
that the flare and quiescence spectra of the ASCA data are similar
to each other. 
We utilize the current spectral-fit results to study the spectral
shapes of the ASCA data in comparison with those of the XMM-Newton
data. 
Figure 7 displays the fit result of the XMM-Newton model of the
quiescent data to the ASCA quiescent spectrum (best-fit
${\rm N_H}$ was $4.9\times10^{20}$ cm$^{-2}$ with
$\chi^2 = 43.0$ for 74 degrees of freedom).
In this fit, all the model parameters, except for the
hydrogen-equivalent column density and the overall normalization,
were fixed to the best-fit values of the quiescent spectrum (R1)
in Table 2.   
We had to set the column density free to adapt the long-term
change of the low-energy efficiency of ASCA SIS.
This result suggests that the ASCA and XMM-Newton quiescent spectra
are consistent with each other. However, the small $\chi^2$ value
also indicate poor statistics of the data.
Thus we try to fit the two models to the ASCA flare spectrum, i.e.
the best-fit models of the XMM-Newton flare (R3) and quiescent (R1)
spectra.
Again, we fixed all the parameters to the best-fit values except
for the hydrogen-equivalent column density and the overall
normalization. We obtained $\chi^2=50.7$ (74 degrees of
freedom, ${\rm N_H}=1.0\times10^{21}$ cm$^{-2}$) for the
quiescent model, and $\chi^2 = 46.4$ (74 degrees of freedom,
${\rm N_H}=3.4\times10^{20}$ cm$^{-2}$) for the flare model.
This means that the ASCA flare spectrum does not have enough
statistics to distinguish the best-fit
models of the flare and the quiescent data of XMM-Newton.
This also holds for the quiescent spectrum of ASCA.
These results indicate that it is difficult to make a meaningful
comparison between the ASCA and XMM-Newton spectra.

To summarize, fresh X-rays are produced at the rising phase of
the flare (R2), but their energy distribution is not much
different from that of the quiescent X-rays.  A spectral
hardening occurred evidently as the flare develops (R3-R4).
There is a spectral component little affected by the flare.

\subsection{Diagnostic of the Energy Spectrum of the Pulsed Emission}

We next explore an energy-resolved pulse profiles to obtain spectral
information for pulsed X-rays.
For this, we select 4 different energy bands taking into account both
data statistics and temperature values in Table~2, i.e., 0.15 $-$ 0.6~keV,
0.6 $-$ 1.2~keV, 1.2 $-$ 3.0~keV, and 3.0 $-$ 10.0~keV.
Figure~8 displays the energy-resolved pulse profiles represented by crosses.
The solid curves represent the best-fit model functions.
Figure~8 clearly reveals that pulsed X-rays exist throughout the
whole energy band.
It is found that all the profiles also have a single broad peak over
one cycle of period.

Subsequently, we fit the pulse profiles using the sinusoidal model as we did
in \S 4 and obtain the pulse-parameter values together with the unpulsed
flux listed in Table~3.
The values indicate that pulsed flux increases more or less evenly
throughout the whole energy band with the increase of total pulsed flux
(see Table~1).
We also calculate PHA ratios shown in Figure~9 between pulsed and unpulsed
fluxes, and find that the ratios are not very simple.
This figure suggests that the spectra of pulsed X-rays are also variable
during the flare and are likely to have multi-temperature components similar
to the spectra of unpulsed X-rays.

\section{Discussion}
\subsection{The Flaring Site}

A flare is observed simultaneously in X-ray and UV during which X-ray
varies together with UV with no significant time lag. 
This correlation suggests that their flaring sites are very close
to each other and have a similar size.
The light curve for the present flare is different from the curves shown
in typical stellar magnetic flares in the sense that the time profile
(especially a flat topped profile for R3), the timescale of flux
variation, and the presence of flickerings \citep[see e.g.,][]{pal90}.
Therefore, the companion can be naturally ruled out from the candidate
of the flaring site.
One may speculate that the best interpretation of the light curve is the
superposition of two different types of flare (R3 \& R4).
However, it is unlikely because the spectra of R3 \& R4 are very similar
to each other (\S 5.2), and the pulsed fluxes change with the flare 
consistently (\S 4).

As mentioned in \S 3.1, flickerings tend to occur more frequently
during the flare. Because the flare may be caused by the enhanced
mass-accretion toward the white dwarf, the accretion flow may have
inhomogeneity such as blobs. We consider the flickerings may be
generated near the magnetosphere through the interactions between
the ambient matter and the inflow of the blobs or the interactions
among the blobs.
In \S 4, we found that both pulsed and unpulsed X-ray fluxes
change with the flare.
The increase of the unpulsed flux itself implies that most of the flare
emission should be generated from outside the magnetosphere.
At the same time, the increase of the pulsed flux (or pulse amplitude)
is an indication that an accreting matter is enhanced onto the white
dwarf surface during the flare.
To satisfy all these observed characteristics, the flare should occur at
a region very close to the magnetosphere of the white dwarf.
This inference agrees with the idea of \citet{era96}.\footnote{Through
the analysis of UV data, they confined the flaring site to a region
somewhere between the primary and the companion star (they suspected
a shock-heated blobs as the origin of the flare).} 

Recently, \citet{ski03} and \citet{pea03} presented an optical light
curve\footnote{We correct the orbital phase coverage of their light
curve $\Phi_{\rm orb} = 0.52 - 0.54$ to be around the phase 0.3,
whose coverage is coincident with the range of the R3 segment in Figure~1.}
and suggested that (small) optical flare\footnote{The amplitude and
timescale of flux variation are comparable to those of the flickerings
(\S 3.1).} of AE Aqr arise from a region outside the binary system
through collisions of blobs expelled from the rotating magnetosphere.
The flaring mechanism looks attractive but the flaring site they suggested
is inconsistent with our results.
If a flare originates from a region far away from the magnetosphere by
collisions, it may be difficult to exert influence on the pulsed and
unpulsed X-rays simultaneously. However, as seen in Figure~4, pulsed and
unpulsed X-rays are affected by the flare, i.e., both the fluxes
clearly increase during the flare.
Their suggestion may be correct in general if X-ray and UV/optical
flares occur independently at different places, but such a possibility is
ruled out by our data that show no significant time lag between X-ray and UV.
The upper limit of the time lag $\sim 1$~min is too short for expelled
blobs to escape the Roche-lobe of the white dwarf.
One may speculate reprocessing of X-rays as an alternative explanation but
this is unlikely because the UV/optical luminosity is larger than the X-ray
luminosity.
We therefore conclude that the most plausible site of the flare is a region
very close to the magnetosphere of the white dwarf.

\subsection{The Mass Accretion Problem}

As showed in \S 4, the unpulsed X-ray flux varies by a factor of 5
during the observation and its time profile is almost the
same as the light curve.
This suggests that the transferred matter or the incoming flow does not
accrete mostly onto the white dwarf surface.
Although there is little agreement on the magnetic field strength of
the white dwarf, the inferred magnetospheric radii exceed a co-rotation
radius by one order of magnitude \citep[e.g.,][]{wyn97, ikh98, nor04}.
Under the circumstance, the behavior of the unpulsed flux can be
understood as the effect of a propeller since the incoming flow 
from the companion can be expelled by the rotating magnetosphere.

On the other hand, the presence of the pulsed X-ray strongly indicates
that an X-ray emitting site should exist at the inner magnetosphere of
the white dwarf and the site should be occulted partly and periodically
by the rotating white dwarf.
In general, the increase of pulsed flux for the flare can easily be
interpreted due to an increase of the mass accretion rate onto the white 
dwarf surface if a propeller activity is not very strong at a high 
magnetic latitude \citep[e.g.,][]{choi00}.
Alternatively, as argued by \citet{ikh01}, if the white dwarf is under the
condition of a $\lq$supersonic propeller' so that an accretion does not
occur completely, then the increase may be a result of some parts of
outer magnetosphere compressing deep into the magnetosphere by compact
blobs \citep[see, e.g.,][]{pea03}.

However, in the latter case, compact blobs would penetrate only down to the
co-rotation radius $r_{\rm co} \equiv (GM_{\rm wd}/\Omega^2)^{1/3}$ =
$(GM_{\rm wd} P_{\rm spin}^2/4\pi^2)^{1/3}$ $\approx 1.4 \times 10^9$ cm.
Accordingly, even if the blobs emit X-rays, it is difficult to expect 
a periodic pulsation from the enhanced radiation because the co-rotation
radius is larger than the radius of the white dwarf so that the blobs
cannot be occulted periodically by the rotating white dwarf.
An additional X-ray source is therefore unavoidable for this case to 
interpret the increase of pulsed flux of the flare.
This additional source should be located temporarily at somewhere
between the white dwarf and the co-rotation radius.
In this sense, the supersonic propeller treatment of AE Aqr and the
$\lq$pulsar-like' white dwarf model suggested by \citet{ikh98}
seem unlikely, since they do not give us a self-consistent explanation for
the increase of pulse parameters during the flare and even for the pulsing
X-ray's behavior of the quiescence.
On the other hand, the former case that a propeller activity is not
too strong does not require such an additional X-ray source.
In this case, if the companion transfers more matter toward the 
magnetosphere, which is the cause of the flare, an accretion could be
enhanced onto the white dwarf surface.

As we found in \S 4, the pulsed flux or the pulse amplitude varies even 
for the quiescence as well as for the flare, which is considered as
a direct evidence of the accretion onto the white dwarf.
This strongly indicates that a propeller action in AE Aqr is actually
not too strong so that the pulsed flux (or pulse amplitude) can easily be
affected by the amount of the accreting matter.
As illustrated in Figure~8 and Figure~9, pulsed X-rays exist through the
entire energy range 0.15 - 10 keV and their spectra are likely to have
multi-temperature components similar to the unpulsed X-ray spectra.
The spectral information of the pulsing X-rays obtained for the
first time in this study implies that the accreting matter is
partly responsible for the generation of the unpulsed X-rays.
It also indicates that the shock of the accreting matter may be
rather moderate compared to those of other magnetic CVs,
because the plasma temperature, a few keV, is low compared to
that of other magnetic CVs, sometimes as high as a few tens keV.
From these analyses and the discussions above, we reach a conclusion that
a small amount of the Roche-lobe overflowed matter accretes onto the white
dwarf (i.e. magnetic pole area) and this matter is responsible for the
X-ray pulsations.
This conclusion supports the idea of Choi et al. (1999), i.e.,
matter gradually drifts toward the pole area without suffering such
a strong shock.
It is therefore natural to explain that the variation of the pulsed flux
is related to the amount of the accreting matter.

It may be worthwhile to mention that there is a spectral component little
affected by the flare (see \S 5.2). For example, the lowest-temperature
values in Table~2 are almost constant during the observation. Moreover,
its normalization values do not show a large variation too (see the 
attached uncertainties).
Its luminosity corresponds to $1 \times 10^{30}$ ergs s$^{-1}$
(0.15 - 10 keV) for the assumed distance of 100 pc.
This result may indicate that the spectral component originates from
a coronal emission of the companion.

\subsection{Flare Properties}

As we noted in \S 3.2, the morphology of the XMM-Newton flare is different 
from that of the ASCA flare in the sense that the overall pattern of
flux variation, the timescale of variation, and the duration including
rise and decay.
Through a visual inspection of the ASCA flare , we roughly estimate that
it has a rise and decay timescales of $\sim 80$ min and $\sim 130$ min,
respectively.
These timescales are much longer than the XMM-Newton flare (see \S 3.1).
We consider that this morphological difference can be caused by the change of
an expelled matter properties, e.g., trajectory, velocity, number of blobs 
and their density etc.

Another interesting point is that Choi et al. (1999) found no significant
variation of pulse amplitude, but we found the amplitude to change by
$\sim$ 50\% from R1 to R4.
From the reanalysis of the ASCA data, we estimate an upper limit of the
variation to be $\sim$ 35\% (we here apply the sinusoidal model, i.e.,
a sine curve plus a constant, to the ASCA pulse profiles as we did in \S 4).
This upper limit indicates that the difference of the amplitude variation 
between ASCA and XMM-Newton flare is marginal.
According to the results of \citet{osb95}, pulse amplitude tends to increase
during flares.
Moreover, we see that there is a hint of the variation during a quiescence
although the uncertainty of their amplitude is quite large.
Therefore, it seems that the amplitude variation of the present study is
a common property.

As we found in \S 5.1 and \S 5.2, a spectral hardening occurs evidently
as the flare advances. That is, the highest-temperature value of R3
spectrum in Table~2, $kT_4$ = 6.6~keV, is twice the temperature of the
quiescence (3.2 keV).
It is however not clear whether such a spectral hardening is temporary,
because of the poor statistics of the ASCA data.

\section{Summary}

Our main results can be summarized as follows.

\begin{enumerate}

\item A flare is observed for AE Aqr with XMM-Newton, during which X-ray 
varies together with UV with no significant time lag. An upper limit of
the lag is estimated to be $\sim 1$~min.
This is the first result to show a close correlation of flux variation
for the X-ray and UV flares of AE Aqr.

\item No distinct indication of an eclipse, e.g., a rapid decrease and
increase of flux around the orbital phase 0.0, is found from the X-ray
and UV light curves. This confirms that AE Aqr is a non-eclipsing binary
system.

\item The X-ray pulse profiles we obtained exhibit a single
broad peak over one cycle of period, confirming previous results. 
We do not find any hints of an interpulse or a secondary peak which
appears in optical/UV pulse profiles.
All the energy-resolved pulse profiles show a single broad peak too.

\item An X-ray light curve variation is dominated by unpulsed X-rays.
This confirms that most of the transferred matter from the companion 
does not accrete onto the white dwarf surface and suggests that a propeller
is operating in this binary system.
Nonetheless, there is a clear evidence of accretion: during the
quiescence the pulsed flux varies with the total X-ray flux throughout 
the entire energy range of 0.15 - 10 keV.
This is the new result of this study and suggests that the propeller
activity is not too strong as argued by \citet{ikh01}.

\item Fresh X-rays are produced at the beginning of the flare but their
energy distribution is not much different from the distribution of the
quiescence X-rays, whereas a spectral hardening occurs clearly as the 
flare evolves.
The spectra of the pulsed X-rays also vary during the flare and are likely 
to have multi-temperature components similar to the spectra of the unpulsed 
X-rays.
There is a spectral component which is little affected by the flare, implying
that the component originates from a coronal emission of the companion.
These spectral analysis results are reported for the first time in this
study.

\end{enumerate}

\acknowledgments

We are grateful to Drs. Nazar R. Ikhsanov, Heon-Young Chang, and Myeong-Gu Park
for careful reading of manuscript and for useful discussions and comments.
We would like to express our gratitude to the anonymous referee for useful
comments which improve our original manuscript much.
We would also like to express our thanks to Dr. Michael Watson for his effort
in making and planning the XMM-Newton observation.
C.S.C. is grateful to the Korea Science \& Engineering Foundation for support
in part by the grant of the basic research program R01-2004-000-1005-0.


\clearpage

\begin{table}
\begin{small}
\caption{Pulse Parameters and Unpulsed flux of each (sub)segment}
\begin{tabular}{lccccc}  \tableline \tableline
Segment & Orbital Phase & Unpulsed Flux    &  Pulse Amplitude  & Pulse Fraction &
$\chi^2$\\
       &               & (counts s$^{-1}$) & (counts s$^{-1}$) & (\%)
& (d.o.f = 4) \\
\tableline
R1     & 0.060 $-$ 0.228 & 1.84$\pm$0.02   & 0.68$\pm$0.04   & 18$\pm$1
& 9.1\\
\tableline
        &     &     & R1 Subsegments &    &    \\
\tableline
R1-1   & 0.060 $-$ 0.116 & 1.56$\pm$0.02   & 0.51$\pm$0.06   & 16$\pm$2
& 2.9\\
R1-2   & 0.116 $-$ 0.172 & 1.80$\pm$0.02   & 0.65$\pm$0.07   & 18$\pm$2
& 1.6\\
R1-3   & 0.172 $-$ 0.228 & 2.11$\pm$0.03   & 0.89$\pm$0.06   & 21$\pm$1
& 9.6\\
\tableline
R2     & 0.228 $-$ 0.275 & 3.50$\pm$0.05   & 0.96$\pm$0.09    & 14$\pm$1
& 1.8\\
\tableline
        &     &     & R2 Subsegments &    &    \\
\tableline
R2-1   & 0.228 $-$ 0.252 & 2.84$\pm$0.06   & 0.92$\pm$0.10     & 16$\pm$2
& 2.7\\
R2-2   & 0.252 $-$ 0.275 & 4.01$\pm$0.08   & 1.00$\pm$0.20     & 12$\pm$2
& 3.8\\
\tableline
R3     & 0.275 $-$ 0.387 & 5.93$\pm$0.04   & 0.98$\pm$0.08   & 8.3$\pm$0.1
& 5.4\\
\tableline
        &     &     & R3 Subsegments &    &    \\
\tableline
R3-1   & 0.275 $-$ 0.331 & 5.70$\pm$0.06   & 0.96$\pm$0.11   & 8.4$\pm$0.1
& 4.7\\
R3-2   & 0.331 $-$ 0.387 & 6.14$\pm$0.06   & 0.99$\pm$0.11   & 8.1$\pm$0.1
& 5.9\\
\tableline
R4     & 0.387 $-$ 0.457 & 6.21$\pm$0.05   & 0.95$\pm$0.09   & 7.6$\pm$0.1
& 4.0\\
\tableline
        &     &     & R4 Subsegments &    &    \\
\tableline
R4-1   & 0.387 $-$ 0.419 & 8.24$\pm$0.09   & 0.98$\pm$0.18     & 5.9$\pm$1.0
& 4.9\\
R4-2   & 0.419 $-$ 0.451 & 5.01$\pm$0.07   & 1.04$\pm$0.13     & 10.4$\pm$1.3
& 3.0\\
\tableline
\tablecomments{
A model of a sine curve plus a constant is fitted to the pulse profiles of
one cycle data, $A \sin (2\pi [\theta - \theta_0]) + C_0$. 
The parameters in the table are defined as follow:
an unpulsed flux is $C_0 - A$, a pulse amplitude is $2A$, and a pulse
fraction is $A/C_0$ (the ratio of the pulsed flux ($A$) to the total
(pulsed + unpulsed) flux).
All the attached errors are at 90\% confidence level.
}
\end{tabular}
\end{small}
\end{table}
\clearpage

\begin{table}
\begin{small}
\caption{Best-Fit Spectral Parameters of each Segment}
\begin{tabular}{lcccc}  \tableline \tableline
\multicolumn{1}{c}{Parameter} &
\multicolumn{4}{c}{Best-Fit Values} \\ \cline{2-5}
 & R1 & R2 & R3 & R4\\
\tableline
$kT_1$ (keV)         & 0.15$^{+0.02}_{-0.02}$ & 0.16$^{+0.04}_{-0.03}$
                     & 0.17$^{+0.02}_{-0.04}$ & 0.17$^{+0.03}_{-0.03}$\\
Norm$_1$ (10$^{-3}$) & 0.4$^{+0.1}_{-0.1}$ & 0.6$^{+0.3}_{-0.3}$
                     & 0.3$^{+0.3}_{-0.1}$ & 0.4$^{+0.2}_{-0.1}$\\
$kT_2$ (keV)         & 0.58$^{+0.03}_{-0.05}$ & 0.56$^{+0.04}_{-0.05}$
                     & 0.63$^{+0.02}_{-0.02}$ & 0.63$^{+0.02}_{-0.02}$\\
Norm$_2$ (10$^{-3}$) & 1.3$^{+0.2}_{-0.2}$ & 2.4$^{+0.4}_{-0.2}$
                     & 4.5$^{+0.3}_{-0.2}$ & 4.7$^{+0.3}_{-0.3}$\\
$kT_3$ (keV)         & 1.1$^{+0.3}_{-0.1}$ & 1.1$^{+0.3}_{-0.1}$
                     & 1.40$^{+0.11}_{-0.08}$ & 1.4$^{+0.2}_{-0.1}$\\
Norm$_3$ (10$^{-3}$) & 0.9$^{+0.2}_{-0.2}$ & 1.9$^{+0.4}_{-0.5}$
                     & 4.1$^{+0.1}_{-0.2}$ & 3.9$^{+1.2}_{-0.8}$\\
$kT_4$ (keV)         & 3.2$^{+0.8}_{-0.3}$ & 3.3$^{+1.0}_{-0.4}$
                     & 6.6$^{+0.9}_{-0.7}$ & 5.6$^{+1.4}_{-0.7}$\\
Norm$_4$ (10$^{-3}$) & 2.1$^{+0.2}_{-0.5}$ & 4.1$^{+0.5}_{-1.2}$
                     & 7.4$^{+0.5}_{-0.6}$ & 8.4$^{+1.0}_{-1.4}$\\
Ratio$^a$            & 1.03$^{+0.02}_{-0.02}$ & 1.01$^{+0.02}_{-0.03}$
                     & 0.95$^{+0.02}_{-0.01}$ & 0.98$^{+0.02}_{-0.01}$\\
N$_{\rm H}$ ($10^{20}$ cm$^{-2}$) & 0.8$^{+0.5}_{-0.5}$ & 1.7$^{+0.9}_{-0.6}$
                                  & 1.7$^{+0.5}_{-0.3}$ & 1.8$^{+0.4}_{-0.3}$\\
$\chi^2$(d.o.f.)     & 767(697) & 675(697) & 908(697) & 846(697) \\
\tableline
 & \multicolumn{4}{c}{Absorbed Flux (10$^{-12}$ ergs cm$^{-2}$ s$^{-1}$)} \\
\tableline
0.15 $-$ 0.60 keV  & 1.85 & 2.62 & 3.75 & 4.03 \\
0.60 $-$ 1.20 keV  & 2.94 & 5.43 & 8.89 & 9.50 \\
1.20 $-$ 3.00 keV  & 1.74 & 3.37 & 6.80 & 7.38 \\
3.00 $-$ 10.0 keV  & 1.07 & 2.09 & 6.89 & 7.12 \\
\tableline
\tablecomments{
Normalization is defined as ($10^{-14}/4\pi D^2$) $\int n_e n_H dV$,
where $D$ (cm) is the distance to the source.
Elemental abundances are fixed to those obtained by \citet{itoh04}: N = 3.51,
O = 0.74, Ne = 0.43, Mg = 0.70, Si = 0.81, S = 0.73, Ar = 0.21, Ca = 0.19,
Fe = 0.47, and Ni = 1.27.
We adopt a VMEKAL model in XSPEC v11.2.0 to fit both PN and MOS spectra
simultaneously, and assume that all the spectral components have the same
elemental abundances and the hydrogen equivalent column density (N$_H$).
The background taken from an annular source free regions are subtracted from
the MOS1 and MOS2 spectra.
The errors are at 90\% confidence level.
}
\tablenotetext{a}{Ratio of continuum normalization between PN and MOS data,
where MOS normalization is fixed to 1.0 and PN normalization is allowed
to be free.}
\end{tabular}
\end{small}
\end{table}

\clearpage

\begin{table}
\begin{small}
\caption{Energy-Resolved Pulse Parameters}
\begin{tabular}{ccccc}  \tableline \tableline
Energy Band  & Unpulsed Flux  & Pulse Amplitude & Pulse Fraction &
$\chi^2$ \\
(keV)        & (counts s$^{-1}$) & (counts s$^{-1}$) & (\%) &
(d.o.f. = 4)\\
\tableline
             &        &   R1 Segment &        &   \\
\tableline
0.15 - 0.60  & 0.57$\pm$0.01 & 0.22$\pm$0.02 & 19$\pm$2 & 1.56\\
0.60 - 1.20  & 0.88$\pm$0.02 & 0.26$\pm$0.03 & 15$\pm$2 & 6.87\\
1.20 - 3.00  & 0.33$\pm$0.01 & 0.17$\pm$0.02 & 26$\pm$3 & 13.60\\
3.00 - 10.0  & 0.072$\pm$0.005 & 0.03$\pm$0.01 & 21$\pm$7 & 4.50\\
\tableline
             &        &   R2 Segment &        &   \\
\tableline
0.15 - 0.60  & 0.95$\pm$0.03 & 0.32$\pm$0.05 & 17$\pm$3 & 3.27\\
0.60 - 1.20  & 1.72$\pm$0.03 & 0.36$\pm$0.06 & 10$\pm$2 & 3.06\\
1.20 - 3.00  & 0.69$\pm$0.02 & 0.22$\pm$0.04 & 16$\pm$3 & 7.10\\
3.00 - 10.0  & 0.14$\pm$0.01 & 0.05$\pm$0.02 & 18$\pm$7 & 0.85\\
\tableline
             &        &   R3 Segment &        &   \\
\tableline
0.15 - 0.60  & 1.44$\pm$0.02 & 0.29$\pm$0.04 & 10$\pm$1 & 5.86\\
0.60 - 1.20  & 2.81$\pm$0.02 & 0.39$\pm$0.05 & 6.9$\pm$0.1 & 4.89\\
1.20 - 3.00  & 1.32$\pm$0.02 & 0.28$\pm$0.03 & 11$\pm$1 & 2.68\\
3.00 - 10.0  & 0.36$\pm$0.01 & 0.04$\pm$0.02 & 5.6$\pm$2.8 & 1.34\\
\tableline
             &        &   R4 Segment &        &   \\
\tableline
0.15 - 0.60  & 1.55$\pm$0.03 & 0.21$\pm$0.05 & 7$\pm$2 & 2.44\\
0.60 - 1.20  & 2.90$\pm$0.04 & 0.41$\pm$0.07 & 7$\pm$1 & 9.78\\
1.20 - 3.00  & 1.42$\pm$0.03 & 0.29$\pm$0.05 & 10$\pm$2 & 5.58\\
3.00 - 10.0  & 0.37$\pm$0.02 & 0.05$\pm$0.02 & 7$\pm$3 & 6.10\\
\tableline

\tablecomments{The details are the same as Table~1.}
\end{tabular}
\end{small}
\end{table}

\clearpage

\begin{figure}
\includegraphics[angle=0,scale=0.80]{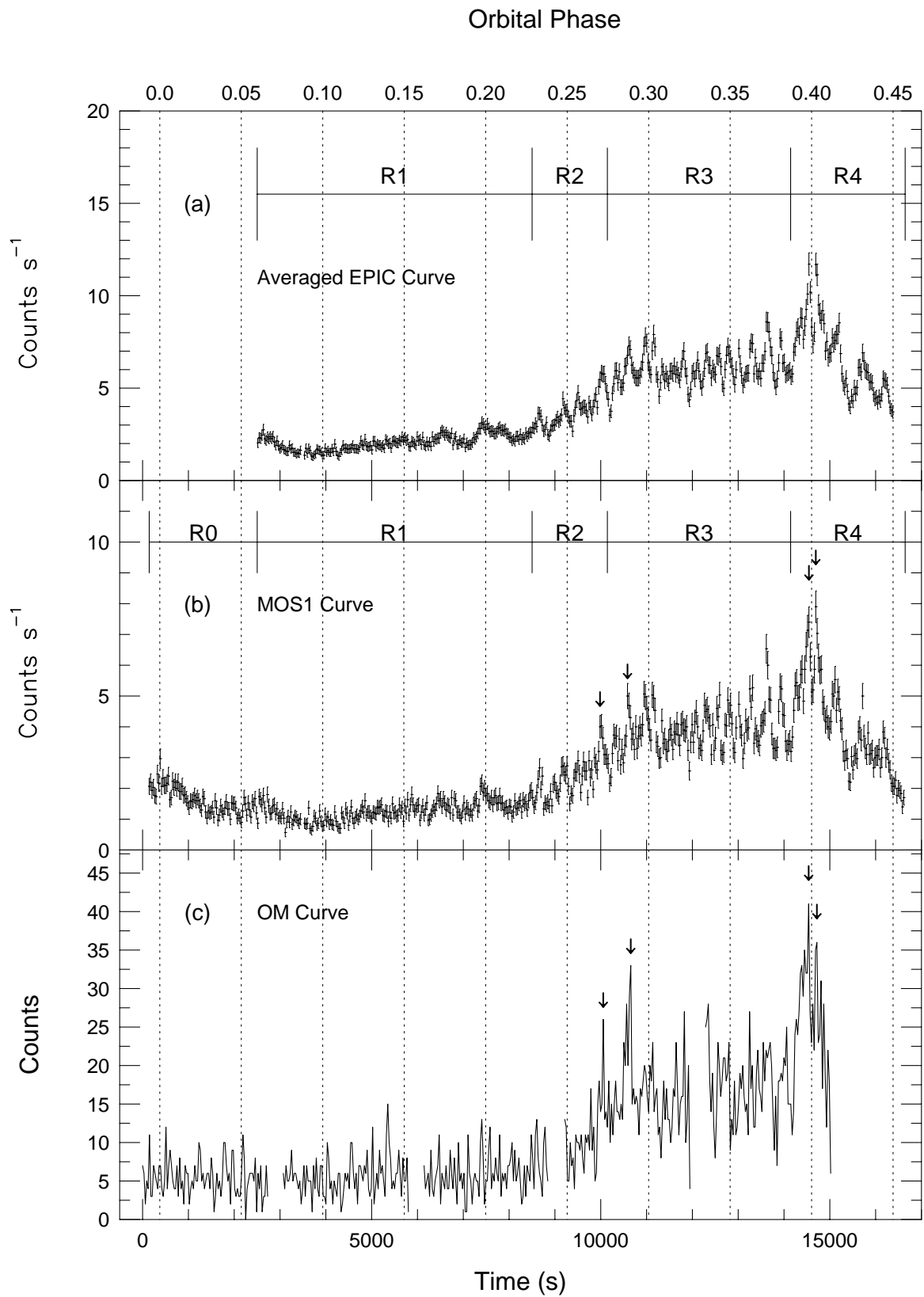}
\caption{
XMM-Newton light curve of AE Aqr: (a) an averaged EPIC curve,
(b) MOS1 curve, and (c) OM curve.
The vertical short-dashed lines represent orbital phases.
All the data were accumulated in 30~s intervals and background which is
small was not subtracted.
In this plot, the time 0 corresponds to MJD 52220.9621 in terrestrial
dynamical time.
}
\end{figure}
\clearpage

\begin{figure}
\includegraphics[angle=270,scale=0.8]{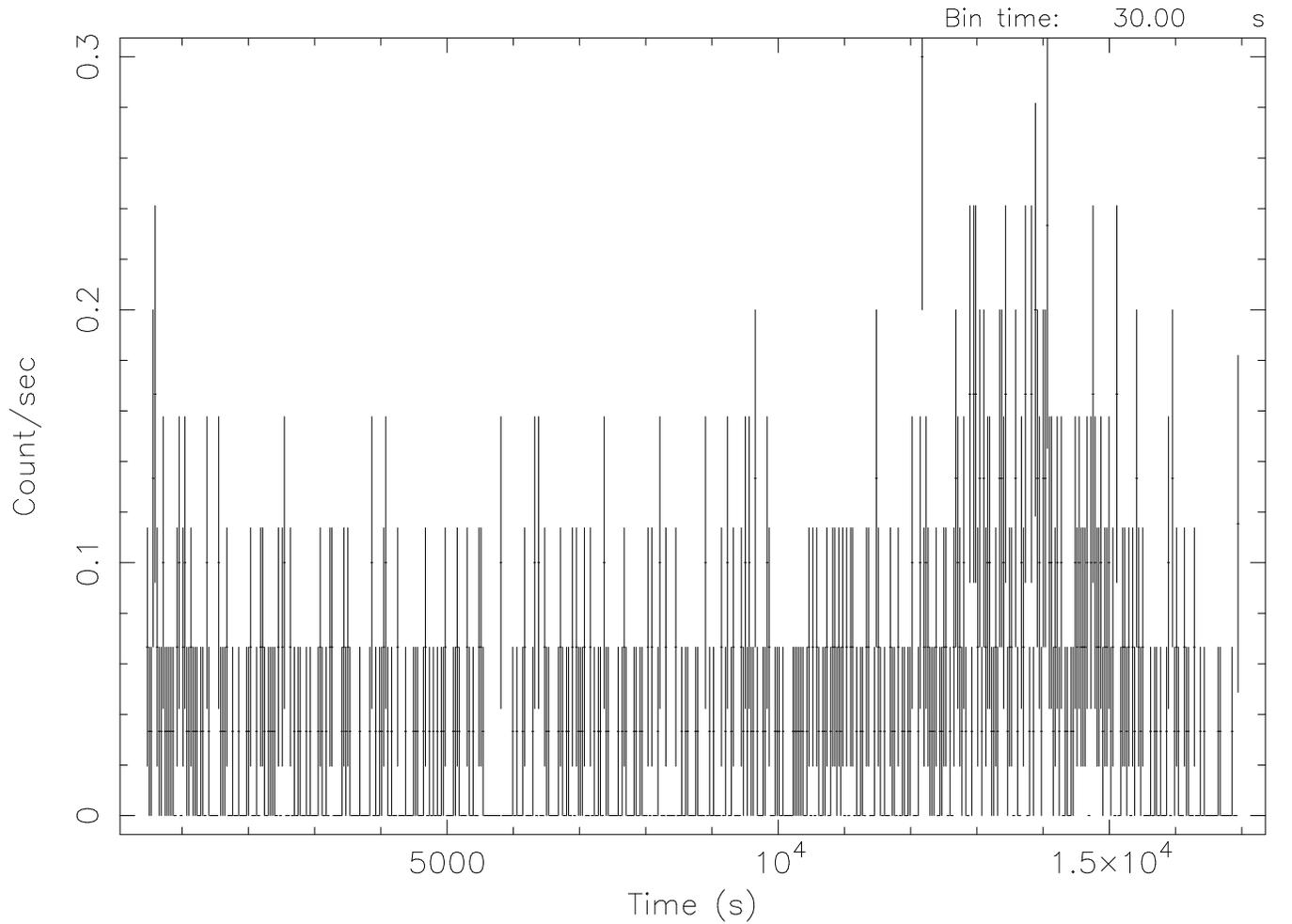}
\caption{
MOS1 light curve extracted from an annular photon deficit region (e.g.,
$\lq$in annulus (27710, 27014, 2740.5, 3405)' in SAS/xmmselect).
As we see, the count rate is remarkably small compared with the rate of
Figure~1b.
This means that the flare is not related with a background flare.
}
\end{figure}

\clearpage

\begin{figure}
\includegraphics[angle=0,scale=1.0]{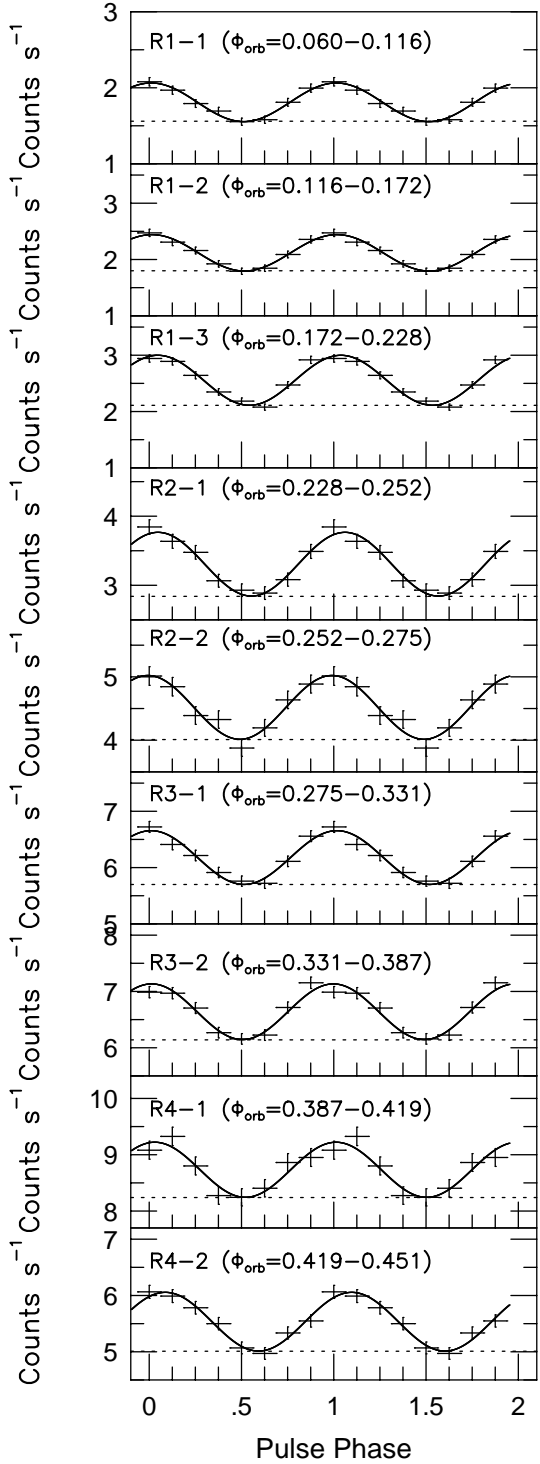}
\caption{
Pulse profile of each smaller segment's data.
A sine curve plus a constant (solid curve) is fitted to
the profiles (crosses).
The horizontal short-dashed lines represent the unpulsed flux levels.
Note that the pulse phase has been repeated over two cycles.
}
\end{figure}

\clearpage

\begin{figure}
\includegraphics[angle=0,scale=0.8]{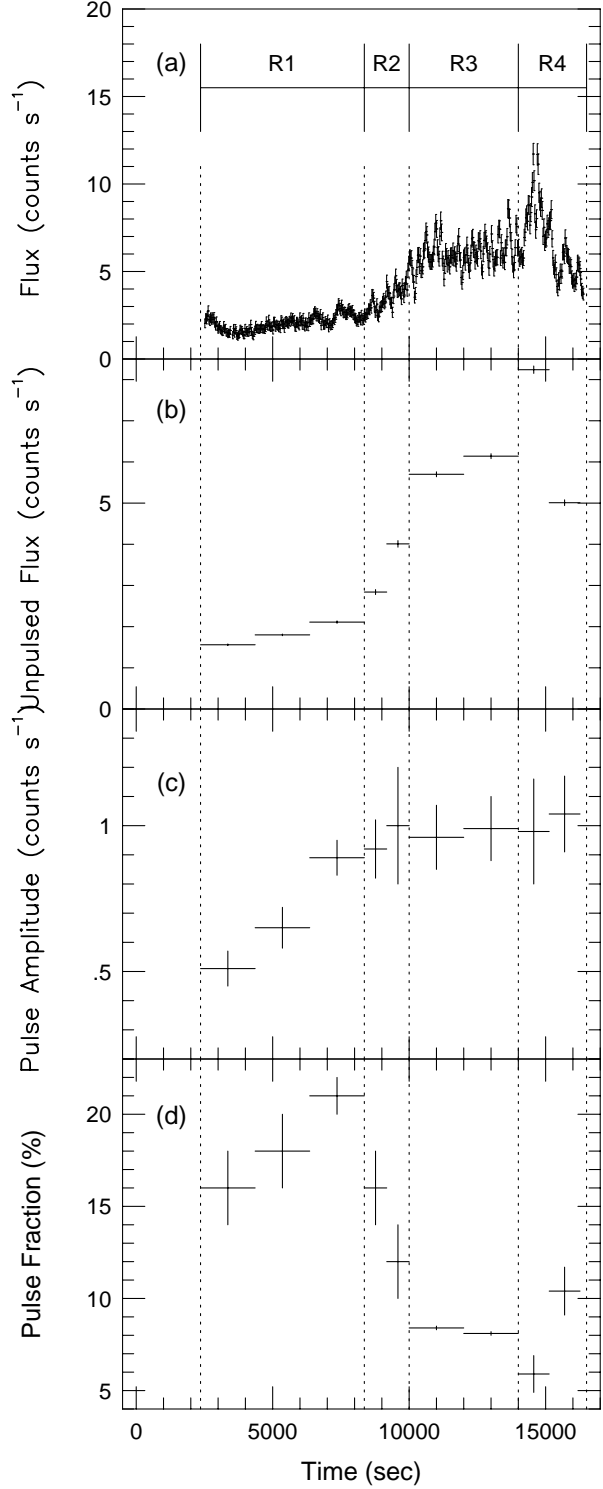}
\caption{
Behaviors of pulse parameters in connection with the light curve
variation.
(a) An averaged EPIC light curve which is the same as the figure~1a.
(b) Unpulsed flux, where the flux represents an average value
detected by EPIC PN and EPIC MOS1 \& MOS2 detectors.
(c) Pulse amplitude. (d) Pulse fraction.
For the definition of pulse parameters, see the table comments 
of Table~1.
All the attached errors are at 90\% confidence level.
}
\end{figure}

\clearpage

\begin{figure}
\includegraphics[angle=0,scale=1.0]{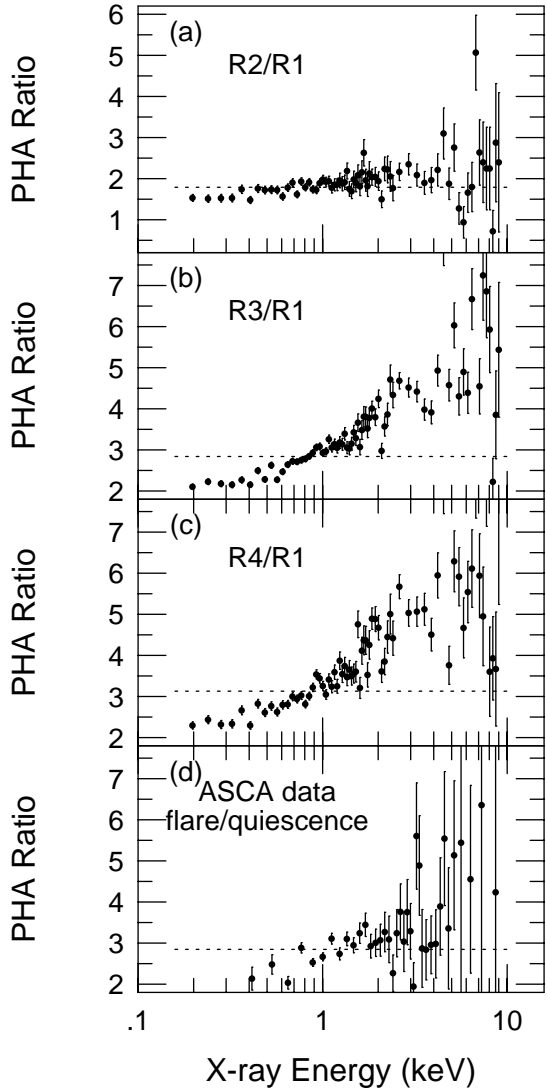}
\caption{
(a) $-$ (c) PHA ratios for R2 $-$ R4 data and (d) for ASCA data.
Each segment's spectrum is divided by the quiescence spectrum (R1) as
a function of X-ray energy.
The horizontal broken lines represent the ratios of average fluxes between
two segments.
}
\end{figure}

\clearpage

\begin{figure}
\includegraphics[angle=-90,scale=0.7]{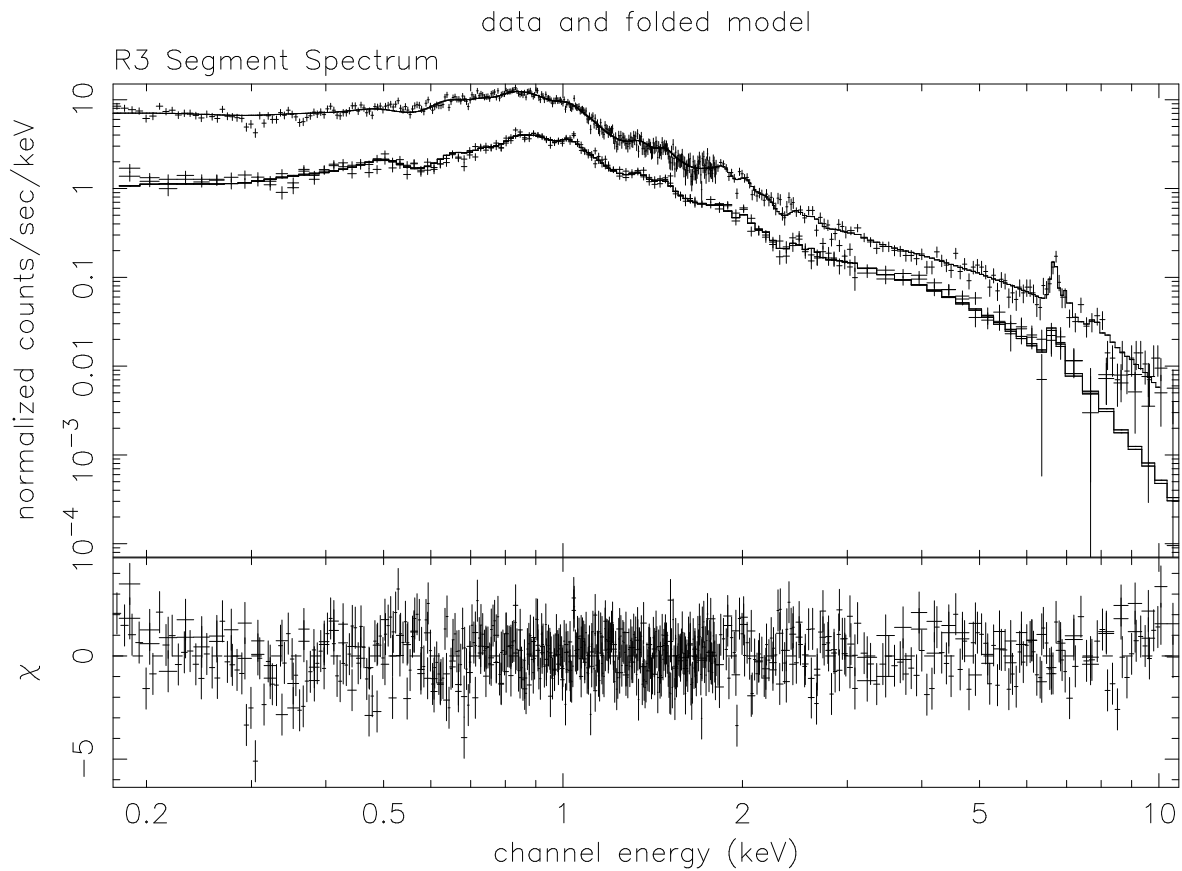}
\caption{
Flare spectra of AE Aqr extracted from the R3 segment.
A function of 4-temperature VMEKAL model is applied to fit both
PN (above) and MOS (below) spectra simultaneously (where elemental
abundances and absorption column densities are linked together).
The histograms represent the best-fit model functions.
A relatively poor fit is mainly ascribed to an insufficient fit
to the data around 0.3 keV (we tried to improve the fit by allowing
a Carbon abundance to be free but failed).
}
\end{figure}

\clearpage

\begin{figure}
\includegraphics[angle=-90,scale=0.7]{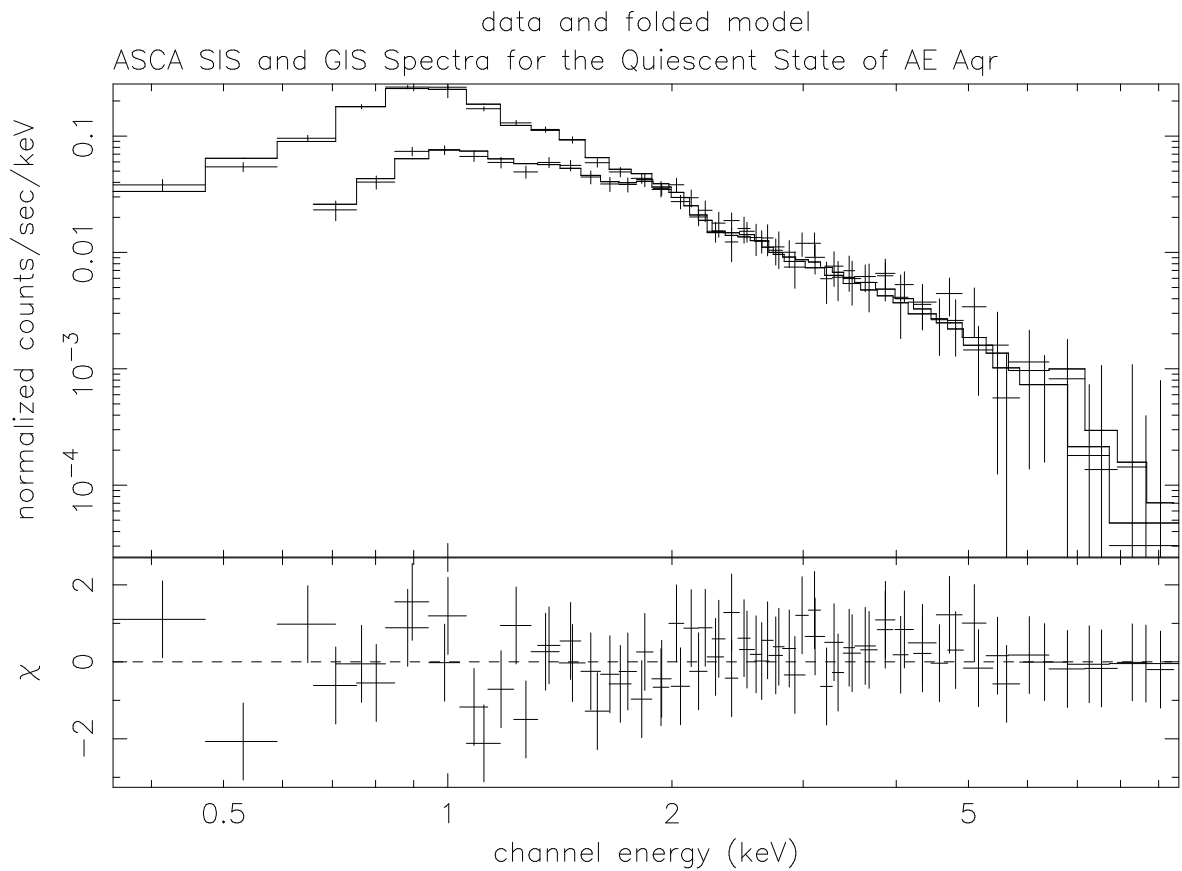}
\caption{
A 4-temperature model fit to the quiescence spectra of ASCA SIS (above)
and GIS (below) data, where the data are the same as that of Choi et al. (1999).
In this fit the parameter values, e.g., temperatures of 4 VMEKAL components,
their relative normalizations, and the elemental abundances, are fixed as 
the best-fit values of the quiescence spectrum (R1) in Table~2.
The overall normalization and the hydrogen-equivalent
column density were optimized to allow the calibration uncertainties.
}
\end{figure}
\clearpage

\begin{figure}
\includegraphics[angle=0,scale=0.8]{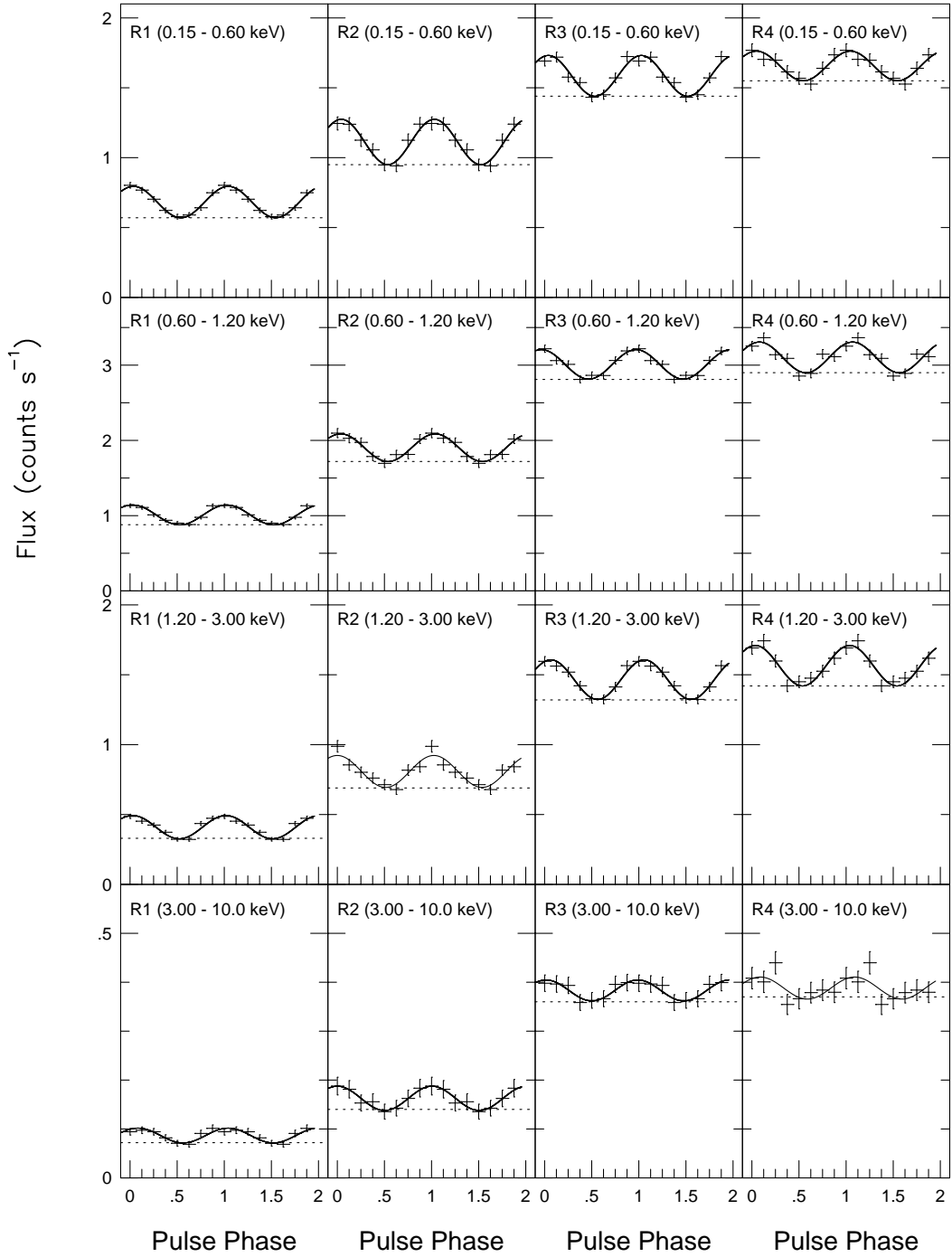}
\caption{
Energy-resolved pulse profiles folded at the period 33.067~s.
Details are the same as Figure~3.
}
\end{figure}

\clearpage

\begin{figure}
\includegraphics[angle=0,scale=1.0]{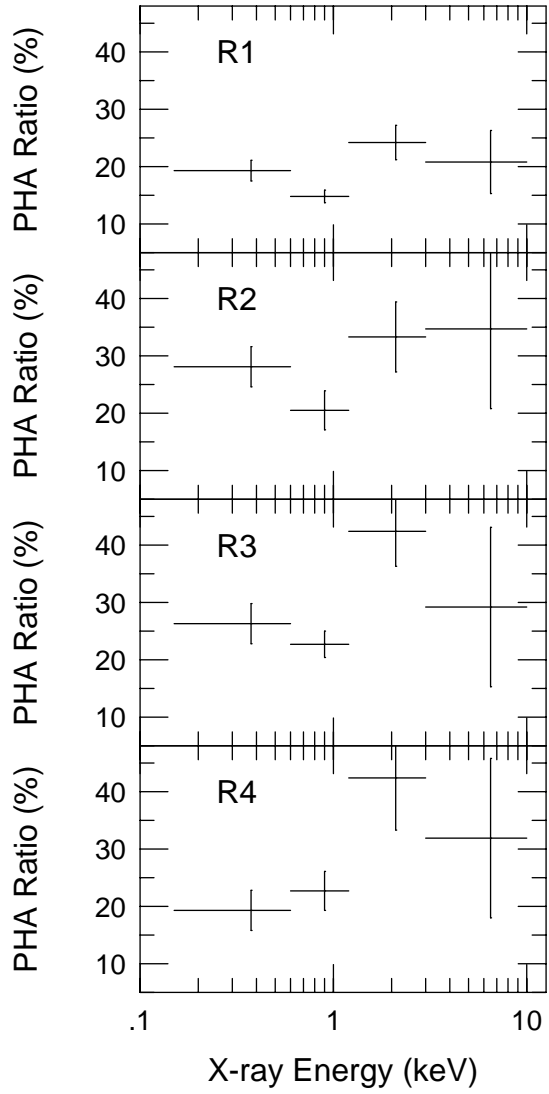}
\caption{
PHA ratios for pulsed X-rays, where each pulsed X-ray spectrum is
divided by the unpulsed X-ray spectrum of the quiescence.
}
\end{figure}



\begin{thebibliography}{}

\bibitem[Abada-Simon et al.(1995)]{aba95}
Abada-Simon, M., Bastian, T. S., Horne, K. et al. 1995,
in Proc. Cape Workshop on Magnetic Cataclysmic Variables, ed.
D.A.H. Buckley \& B. Warner (San Francisco:ASP), 355 

\bibitem[Bastian, Dulk, \& Chanmugam(1988)]{bas88} 
Bastian, T. S., Dulk, G. A., \& Chanmugam, G. 1988, \apj, 324, 431

\bibitem[Casares et al.(1996)]{cas96} 
Casares, J., Mouchet, M., Martinez-Pais, I. G., \& Harlaftis, E. T.
1996, \mnras, 282, 182

\bibitem[Choi \& Dotani(1999)]{cd99}
Choi, C. S., \& Dotani, T. 1999, AN, 320, 348

\bibitem[Choi, Dotani, \& Agrawal(1999)]{choi99}
Choi, C. S., Dotani, T., \& Agrawal, P. C. 1999, \apj, 525, 399

\bibitem[Choi \& Yi(2000)]{choi00} 
Choi, C. S., \& Yi, I. 2000, \apj, 538, 862

\bibitem[Clayton \& Osborne(1995)]{cla95} 
Clayton, K. L., \& Osborne, J. P. 1995, in Proc. Cape Workshop on
Magnetic Cataclysmic Variables, ed. D.A.H. Buckley \& B. Warner
(San Francisco:ASP), 379

\bibitem[den Herder et al.(2001)]{her01}
den Herder, J. W., Brinkman, A. C., Kahn, S. M. et al. 2001, \aap, 365, L7

\bibitem[Eracleous \& Horne(1996)]{era96}
Eracleous, M., \& Horne, K. 1996, \apj, 471, 427

\bibitem[Eracleous et al.(1994)]{era94}
Eracleous, M., Horne, K., Robinson, E. L., Zhang, E. -H., Arsh, T. R., \&
Wood, J. H. 1994, \apj, 433, 313

\bibitem[Ikhsanov(1998)]{ikh98}
Ikhsanov, N. R. 1998, \aap, 338, 521

\bibitem[Ikhsanov(2001)]{ikh01}
Ikhsanov, N. R. 2001, \aap, 374, 1030

\bibitem[Itoh, Ishida, \& Kunieda(2004)]{itoh04}
Itoh, K., Ishida, M., \& Kunieda, H. 2004, astro-ph/0412559

\bibitem[Jansen et al.(2001)]{jan01}
Jansen, F., Lumb, D., Altieri, D. et al. 2001, \aap, L1

\bibitem[Larsson(1996)]{lar96}
Larsson, S. 1996, \aaps, 117, 197

\bibitem[Mason et al.(2001)]{mas01}
Mason, K. O., Breeveld, A., Much, R. et al. 2001, \aap, 365, L36

\bibitem[Norton, Wynn, \& Somerscales(2004)]{nor04}
Norton, A. J., Wynn, G. A., \& Somerscales, R. V. 2004, \apj, 614, 349

\bibitem[Osborne et al.(1995)]{osb95}
Osborne, J. P., Clayton, K. L., O'Donoghue, D., Eracleous, M.,
Horne, K., \& Kanaan, A. 1995, in Proc. Cape Workshop on Magnetic
Cataclysmic Variables, ed. D.A.H. Buckley \& B. Warner (San Francisco:ASP), 368

\bibitem[Pallavicini, Tagliaferri, \& Stella(1990)]{pal90}
Pallavicini, R., Tagliaferri, G., \& Stella, L. 1990, \aap, 228, 403

\bibitem[Patterson(1979)]{pat79}
Patterson, J. 1979, \apj, 234, 978

\bibitem[Pearson, Horne, \& Skidmore(2003)]{pea03}
Pearson, K. J., Horne, K., \& Skidmore, W. 2003, \mnras, 338, 1067

\bibitem[Reinsch \& Beuermann(1994)]{rb94}
Reinsch, K., \& Beuermann, K. 1994, \aap, 282, 493

\bibitem[Skidmore et al.(2003)]{ski03}
Skidmore, W., O'Brien, K., Horne, K., Gomer, R., Oke, J. B, \& Pearson, K. J.
2003, \mnras, 338, 1057

\bibitem[Str\"uder et al.(2001)]{str01}
Str\"uder, L., Briel, U., Dennerl, K. et al. 2001, \aap, 365, L18

\bibitem[Turner et al.(2001)]{tur01}
Turner, M. J. L., Abbey, A., Arnaud, M. et al. 2001, \aap, 365, L27

\bibitem[van Paradijs, Kraakman, \& van Amerongen(1989)]{par89}
van Paradijs, J., Kraakman, H., \& van Amerongen, S., 1989, \aaps, 79, 205

\bibitem[Warner(1995)]{war95}
Warner, B. 1995, Cataclysmic Variable Stars (Cambridge:Cambridge Univ. Press)

\bibitem[Welsh, Horne, \& Gomer(1993)]{wel93}
Welsh, W. F., Horne, K., \& Gomer, R. 1993, \apj, 410, L39

\bibitem[Welsh, Horne, \& Gomer(1998)]{wel98}
Welsh, W. F., Horne, K., \& Gomer, R. 1998, \mnras, 298, 285

\bibitem[Wynn, King, \& Horne(1997)]{wyn97}
Wynn, G. A., King, A. R., \& Horne, K. 1997, \mnras, 286, 436

\end{thebibliography}
\end{document}